\documentclass[twocolumn,aps,amsmath,amssymb,showpacs]{revtex4}
\usepackage{graphicx}
\usepackage{epsfig}

\begin{document}

\title{Additive-multiplicative stochastic models of 
financial mean-reverting processes }
\author{C. Anteneodo}
\email{celia@cbpf.br}
\affiliation{Departamento de F\'{\i}sica, Pontif\'{\i}cia 
Universidade Cat\'olica do Rio de Janeiro, \\
CP 38071, 22452-970, Rio de Janeiro, Brazil}
\author{R. Riera}
\email{rrif@fis.puc-rio.br}
\affiliation{Departamento de F\'{\i}sica, Pontif\'{\i}cia 
Universidade Cat\'olica do Rio de Janeiro, \\
CP 38071, 22452-970, Rio de Janeiro, Brazil}

  
\begin{abstract} 
We investigate a generalized stochastic model with 
the property known as mean reversion, that is, the  
tendency to relax towards a historical reference level. 
Besides this property, the dynamics is driven by 
multiplicative and additive Wiener processes. 
While the former is modulated by the internal behavior of the 
system, the latter is purely exogenous. 
We focus on the stochastic dynamics of volatilities, but our model may also be   
suitable for other financial random variables exhibiting the 
mean reversion property. 
The generalized model contains, as particular cases, 
many early approaches in the literature of volatilities or, more generally, 
of mean-reverting financial processes.   
We analyze the long-time probability density function associated to 
the model defined through a It\^o-Langevin equation.   
We obtain a rich spectrum of shapes for the probability function 
according to the model parameters. 
We show that additive-multiplicative processes provide realistic models 
to describe empirical distributions, for the whole range of data.
\end{abstract} 
 

\maketitle

\section{Introduction}

Accurate statistical description of the stochastic dynamics of stock prices 
is fundamental to investment, option pricing and risk management. 
In particular, a relevant quantity is the volatility of price time series\cite{volatility}, 
that quantifies the propensity of the market to fluctuate. 
Since volatility represents a measure of the risk associated to the fluctuating dynamics of 
prices, it is crucial to develop suitable models to predict 
its complex intermittent behavior. 
There is empirical evidence that it fluctuates following a stochastic 
dynamics subjacent to that of prices, whose dynamics, in turn,  
depends on the time evolving volatility. 
Many approaches are based on that assumption\cite{prices}, although 
others propose the existence of a reciprocal 
feedback between both processes\cite{leverage}. 
 
Our approach builds on the development of a simple Langevin equation to 
characterize the stochastic process of volatility. The equation provides 
an unifying description that generalizes widely discussed models in the literature. 
We analyze the shape of  
the long-time probability density function (PDF) associated 
to the stochastic differential equation that characterizes each 
particular case of the generalized model. 
Most previous results focus on the tails of the PDFs. 
In fact, for stochastic variables, such as volatilities 
presenting fat-tailed PDFs\cite{liu,volat1}, it is specially important 
to reproduce extreme events in a realistic model. 
Now we go a step further and aim to predict the PDFs in the whole range of events. 

One of the main features observed in the dynamics of some financial variables, 
such as volatilities, stock volumes or interest rates, is their tendency to 
permanently relax,  towards a reference level $\theta$, 
a  property known as mean reversion. 
Another feature is the multiplicative market processing of random news, 
whose strength becomes modulated by a function of the stochastic variable itself. 
These two properties are modeled by means of a nonlinear mean-reverting force and nonlinear 
multiplicative noise. They are discussed in detail in Sect. \ref{Sec:props}. 

In Sect.~\ref{Sec:mean_rev}, we discuss the shapes of the PDFs 
that such family of models yields. 
Despite being of a general form, they give rise to PDFs that, 
decay exponentially fast, either above the mode, below it, or both, in 
disagreement with empirical observations. 
For instance, log-normal behavior has been reported for volatility  computed from global data 
of S\&P500\cite{liu}, at intermediate values. 
However, at high values, a power-law behavior, with exponent outside the stable L\'evy range, 
was observed.  
The same analysis performed for individual companies\cite{liu} yields also power-law tails. 
But in that case, the results show a variation slower than log-normal below 
the mode, suggesting a power-law also in the limit of small values. 
The volatility of capitalized 
stocks traded in US equity markets exhibits similar features\cite{micciche}. 
Other variables with mean-reversion,  
such as volume of transactions (number of trades) present akin distributions.  
Power-law tails out of the L\'evy range have been reported 
for the PDFs of normalized NYSE stock volumes\cite{volumes}. 
More recently, studies of normalized volumes, performed over 
high resolution data (1-3 minutes) of NYSE and NASDAQ\cite{vol1} (see also \cite{vol2}), 
display PDFs with power-law behavior both at large and small values. 
We will show that the class of multiplicative processes considered in 
Sect. \ref{Sec:mean_rev}, although general enough,
is not able to reproduce, for any value of its 
parameters, these empirical PDFs in the whole range. 
 
In a realistic model, we must deal with various sources of fluctuations acting upon 
the collective variable. Then, we propose to include a component that is lacking 
to suitably model many real processes, that 
is the presence of fluctuations that act additively, 
besides the multiplicative noise already taken into account. 
The latter originates from the internal correlated behavior of the 
market, representing a sort of endogenous feed-back effect, while additive noise concerns 
fluctuations of purely external origin or random speculative trading. 
Then, in Sect. \ref{Sec:add_mult}, we present a further generalization that 
consists in incorporating an independent additive source of noise.  
Depending on the parameters of the process, the additive or multiplicative contributions 
will play the dominant role. This gives rise to a rich spectrum of PDF shapes,   
in particular, a subclass with two-fold power-law behavior, both above and below the mode,  
providing a general realistic framework for describing the shape of empirical distributions. 
A comparison with experimental results is presented in Sect. \ref{Sec:empirical}. 
Finally, Sect. \ref{Sec:final} contains the main conclusions and general remarks.

\section{Mean reversion and multiplicative fluctuations} 
\label{Sec:props}

The reversion to the mean is one of the basic ingredients to  
describe the dynamics of several stochastic variables of interest in economy.   
It is fundamental since it concerns the behavior around a central value $\theta$ 
and reflects the global market response to deviations from a consensus or 
equilibrium level. 
It depends on monetary unit, market size, degree of risk aversion, etc., 
hence, it is characteristic of each market. 
The aversion to deviations from the mean needs not be linear, specially when large 
deviations are involved. 
Similarly, a nonlinear mechanism due to the cooperative behavior of traders, rules the way 
the market modulates the amplitude of fluctuations (mainly external) 
giving rise to innovations.  

We consider the general class of stochastic differential equations given by
\begin{equation} \label{ILE0}
{\rm d} x \;=\;-\gamma[x-\theta]x^{r-1}{\rm d}t + \mu x^s{\rm d}w \;,
\end{equation}
where, $\theta, \gamma,\mu>0$, $r,s\in\mathbb{R}$, and $w$ is a Wiener process, such that 
$\langle {\rm d}w\rangle=0$ and $\langle ({\rm d}w)^2\rangle=2{\rm d}t$.  
The definition of the stochastic process is completed by the It\^o prescription. 
This class generalizes well-known models employed to describe the dynamics of 
mean-reverting financial variables\cite{juros}.
In particular, some traditional processes for modeling volatilities  or, mainly, 
squared volatilities  are the Hull-White ($r=1,s=1$)\cite{hw} and  
the Heston ($r=1,s=1/2$)\cite{heston} models, the latter  
also known either as Cox-Ingersoll-Ross\cite{cir} or Feller process\cite{feller}. 
The arithmetic ($r=1,s=0$) and  geometric ($r=2,s=1$)
Ornstein-Ulhenbeck processes  are particular cases too. 
Moreover, several other models employed in  the literature of volatilities are related to 
this class\cite{others,micciche}.

Different values of $r$ in Eq. (\ref{ILE0}) represent different possible 
relaxation mechanisms of amplitude $\gamma$,  
determined, amongst other factors, by  constraints, flux of information, stock liquidity  
and risk aversion, which are particular of a given market.  
Notice that the restoring force in Eq. (\ref{ILE0}) corresponds to a confining potential, 
with minimum at $\theta$, for all $r\in\mathbb{R}$. 
The larger $r$, the more attractive the potential for large $x$, but the less 
attractive for vanishing $x$. 
Similarly, different values of $s$ specify the market informational connectivity, 
which conditions the degree of coherent multiplicative behavior.   
Models in the literature typically set $s\geq 0$, meaning that the effective amplitude 
of fluctuations increases with $x$. 
Negative $s$ makes multiplicative fluctuations grow with 
decreasing $x$, thus it mainly reflects a cooperative reaction to quiescence. 
Although it does not seem to reflect a realistic  
steady state of the market, it may occur as a transient, driven by speculative trading.  

The two mechanisms are complementary. 
If $r<0$ the restoring force decreases for increasing $x$ 
above the reference level, in particular, for $r<-1$, the force tends to zero in 
the limit of large $x$. Thus, decreasing $r$ represents markets that, 
become less able to recover the reference level by means of the 
deterministic tendency alone. However,  a strong multiplicative response to large 
fluctuations (positive $s$) could 
still compensate that inability and restore the market historical level.  
Concerning the response to small values, the restoring force diverges at the origin if 
$r<1$, while for $r>1$, it vanishes at $x=0$, meaning that this point becomes an unstable 
equilibrium state. This corresponds to a market indifferent to low levels 
of trading activity. 
Again, this effect can be balanced by the multiplicative contribution  
(with a small value of parameter $s$).

In early works, only very particular values of $(r,s)$ have been considered. 
However, this may be sometimes owed more to reasons of mathematical solvability, 
than to econophysical ones. 
Following the above discussion, $(r,s)$ may be non-universal, 
depending on the particular nature of a market or its agents. 
Therefore, we will not discard any possibility {\em a priori}.

\section{Generalized multiplicative process with mean reversion} 
\label{Sec:mean_rev}

We consider the simple class of stochastic multiplicative differential 
equations given by Eq. (\ref{ILE0}), that generalizes many 
processes usually found in the literature of volatilities. 
We investigate, in this Section, the long-time PDFs that this class of processes yields.  
The Fokker-Planck equation associated to Eq.~(\ref{ILE0}), following  
standard methods\cite{books}, is
\begin{equation} \label{FP0}
\partial_t\rho 
= \gamma \partial_x( [x-\theta]x^{r-1}\rho) + \mu^2 \partial^2_{xx} [x^{2s}\rho] \;.
\end{equation}
Its long-term solution is relevant in connection to the 
assumption that the process can be treated as quasi-stationary. 
In that case the PDF obtained from an actual data series will coincide with the 
stationary solution. 
Considering reflecting boundary conditions at $x=0$ and $x\to\infty$\cite{books}, 
the steady state solution of Eq. (\ref{FP0}) reads: 

\begin{equation} \label{ss0}
\rho(x)=\frac{\rho_o}{x^{2s}}  \exp\left( -\gamma_\mu\Bigl[
\frac{x^{p+1}}{p+1} - \theta\frac{x^{p}}{p} \Bigr]\right),   
\end{equation}  
with $p\equiv r-2s\neq0,-1$,  
where $\rho_o$ is a normalization constant and $\gamma_\mu\equiv\gamma/\mu^2$  
an effective restoring amplitude, such that  $\gamma$ (a parameter associated to order) 
becomes reduced by the amplitude of multiplicative noise (associated to disorder).

\begin{figure}[htb] 
\begin{center} 
\vspace*{1cm}
\includegraphics*[bb=140 450 460 720, width=0.4\textwidth]{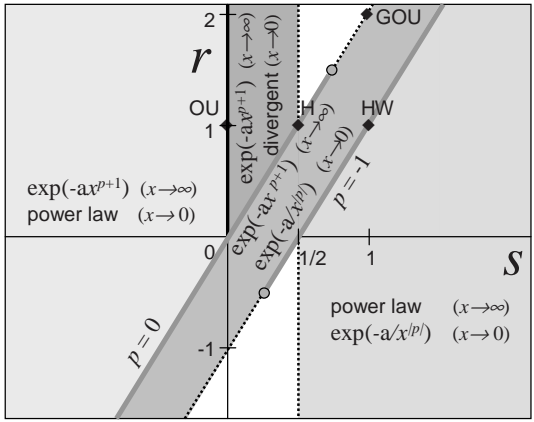} 
\end{center} 
\caption{
Diagram of  the asymptotic behavior of the PDF given by 
Eq. (\ref{ss0}), in ($s,r$)-plane. 
Unshadowed regions and dotted borders identify regions excluded by 
the normalization condition.  
At the positive $r$-axis, the PDF is finite at the origin. 
Tilted lines denote the marginal cases $r=2s$ ($p=0$), with pure exponential tail and 
power-law growth at the origin, and $r=2s-1$ ($p=-1$), with power-law tail and exponential 
of $1/x$ growth at the origin (the threshold points of these lines have coordinates  
$([1+\gamma_\mu\theta]/2,0)$ and $([1-\gamma_\mu]/2,0)$, respectively).
Parameter $a>0$, in the exponential formulas, as well as the power-law exponents, depend on 
model parameters. 
Symbols correspond to the special processes: Hull-White (HW), Heston (H), 
Ornstein-Ulhenbeck (OU) and geometric OU (GOU).   
} 
\label{fig:m} 
\end{figure}

The class of processes described by Eq. (\ref{ss0}) thus generically yields 
asymptotic exponential-like behaviors for small or/and large values. 
As soon as $p+1>0$,  a {\em stretched exponential decay}  is obtained 
for large enough $x$, such that the argument of the exponential is proportional 
to $-x^{p+1}$. 
If $p<0$, a {\em stretched exponential of the inverse argument}  ($-1/x^{|p|}$) 
is obtained for vanishing $x$. 
Therefore, for $p\in(-1,0)$, the PDF presents dominant exponential-like behavior  
both for low and large values, without any restriction on the value of $s$.  
Outside that interval,  
the {\em power law} $x^{-2s}$ in (\ref{ss0}) asymptotically dominates, 
for either small (if $p>0$) or large (if $p<-1$) argument. 
Then, normalization in $[0,\infty)$ restricts the possible values of $s$ 
according to: $s<1/2$ (if $p>0$), $x>1/2$ (if $p<-1$).

In the marginal cases, Eq. (\ref{ss0}) explicitly is: 

~I: For $p\equiv r-2s=-1,$  
\begin{equation} \label{caseI}
\rho(x)=\frac{\rho_o}{x^{2s+\gamma_\mu}}\;
\exp( -\gamma_\mu\theta/x ) \;,
\end{equation} 
with $2s>1-\gamma_\mu$ for nomalizability.

II: For $p\equiv r-2s=0$,    
\begin{equation} \label{caseII}
\rho(x)=\rho_o x^{\gamma_\mu\theta-2s}
\exp( -\gamma_\mu x ) \;,
\end{equation} 
with $2s<\gamma_\mu\theta+1$ for normalization, but 
$2s<\gamma_\mu\theta$ to avoid the divergence at the origin.

Fig.~\ref{fig:m} displays the possible PDF asymptotic shapes in $(s,r)$ space. 
Notice that the $s=0$ axis gives the solution for mean-reverting models with purely additive 
fluctuations. 
Let us analyze some special cases. 
In the trivial case $(s,r)=(0,1)$, corresponding to the Ornstein-Ulhenbeck 
process \cite{books}
\begin{equation} \label{OU}
{\rm d} x \;=\;-\gamma[x-\theta]{\rm d}t + \mu {\rm d}z \;,
\end{equation}
the noisy contribution becomes additive and the 
stationary PDF is Gaussian (truncated at $x=0$). 

Although we are dealing with $\theta>0$, it is worth of mention the case  
$(\theta,s,r)=(0,1,1)$, corresponding to the geometric Brownian process, 
that leads to the {\em log-normal} distribution. 

For type I ($r=2s-1$), notice that the PDF decays as a power law, for large $x$, and 
goes to zero faster than power law, for vanishing $x$. 
The power-law exponent is controlled by $s$ and $\gamma_\mu$,  
that is, all the model parameters, except $\theta$ are involved. 
In the particular case  $r=s=1$, 
one recovers the Hull-White process\cite{hw} 
\begin{equation} \label{HW}
{\rm d} x \;=\;-\gamma[x-\theta] {\rm d}t + \mu x{\rm d}z \;. 
\end{equation}

In case II ($r=2s$), observe that the PDF has opposite behavior: it 
increases at the origin as a power law
and decays exponentially for large $x$. 
All the model parameters, including $\theta$ are buried in the power-law exponent. 
In particular, if $r=2s=1$, one gets the Heston model\cite{heston}
\begin{equation} \label{H}
{\rm d} x \;=\;-\gamma[x-\theta]{\rm d}t + \mu \sqrt{x}{\rm d}z \;. 
\end{equation}  
If $r=2s=2$, the geometric Ornstein-Uhlenbeck process is obtained
\begin{equation} \label{Geo}
{\rm d} x \;=\;-\gamma[x-\theta]x{\rm d}t + \mu x{\rm d}z \;.
\end{equation}
 
Diverse other models proposed in the literature can also be thought as particular instances 
of our generalized model. For example, the one proposed by Miccich\'e et al. \cite{micciche}    
is in correspondence with the Hull-White model (\ref{HW}), with $x$ 
representing volatility $v$, whereas in the latter $x\equiv v^2$. 
Also a family of multiplicative models, studied before in the context of  
a wide spectrum of physical processes\cite{schenzle}, 
belongs to the class here considered, through the transformation $x\to x^\beta$. 
 
Summarizing, from Eqs.~(\ref{ss0})-(\ref{caseII}),
in general, the asymptotic behaviors below and above the mode are tied, such that,  
in a log-log scale, if one flattens the other changes rapidly. 
This explains why models of this class fail to describe empirical volatilities 
in the whole range of observed data, even under the transformation $v^2\mapsto v$.

\section{Generalized model with additive-multiplicative structure}
\label{Sec:add_mult}

We analyze in this section, processes that take into account 
the presence of some additional source of noise. 
Previous works\cite{multi1,multi2,multi3}  
show that additive-multiplicative 
stochastic processes constitute an ubiquitous mechanism leading 
to fat-tailed distributions and correlated sequences. 
This extra noise represents a quite realistic feature, since,  
besides noise modulated by the market, other 
fluctuations may act directly, additively. 
From the stream of news, represented by a noisy signal, some are 
amplified or reduced by cooperative actions, others incorporated unaltered. 
Related ideas has been discussed in Ref. \cite{sornette}. 
Also, a model of financial markets that leads to additive-linear-multiplicative processes 
has been recently proposed\cite{spins}, where the noises are identified with 
the fluctuating environment and fluctuating interaction network, respectively. 
In general, the two white noises are considered uncorrelated. 
However, they may even correspond to identical time-series as soon as 
they are shifted with a time lag  greater than the 
correlation time. In such case, 
the endogenous noise is expected to act with a delay due to 
its very nature of feedback process, whereas, the additive noise 
is incorporated immediately, free of signal processing.

\begin{figure}[hby] 
\begin{center} 
\vspace*{1cm}
\includegraphics*[bb=140 420 465 710, width=0.40\textwidth]{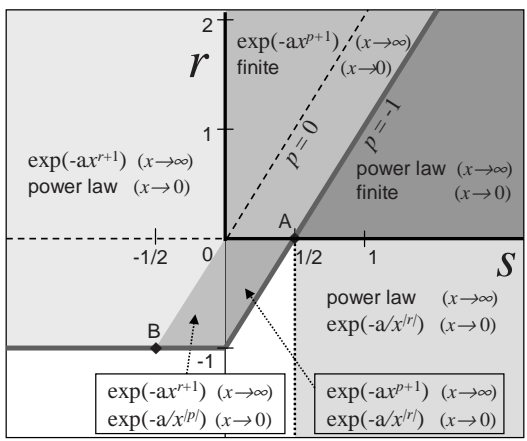} 
\end{center} 
\caption{
Diagram of the asymptotic behavior of the PDF defined by Eq. (\ref{FPss}), 
in $(s,r)$-plane. 
Unshadowed regions and dotted borders are regions excluded by the normalization requirement. 
At both positive semi-axes, the growth at the origin is power law.
On dark gray lines, tails are power law, with the tilted line corresponding 
to $r=2s-1$ ($p=-1$). Dashed lines correspond to pure 
exponential tails, with the tilted line corresponding to 
$r=2s$ ($p=0$).
In the formulas,  $a>0$, as well as the power-law exponents, 
generically depend on model parameters, moreover $p\equiv r-2s$. 
Symbols correspond to the special processes {\bf A} [Eq. (\ref{A})] and 
{\bf B} [Eq. (\ref{B})].
} 
\label{fig:am} 
\end{figure}

By including purely exogenous fluctuations,  in the process defined by Eq. (\ref{ILE0}), 
we obtain the following It\^o-Langevin equation (ILE) 
\begin{equation} \label{ILE}
{\rm d} x \;=\;-\gamma[x-\theta]x^{r-1}{\rm d}t + \mu x^s{\rm d}w + \alpha {\rm d}z \;,
\end{equation}
where $w,z$ are two {\em independent} standard Wiener processes, defined as above,  
and $\mu,\alpha$ their respective amplitudes. 
The corresponding Fokker-Planck equation reads
\begin{equation} \label{FP}
\partial_t\rho 
=  \gamma \partial_x( [x-\theta]x^{r-1}\rho) + \partial_{xx}( [\mu^2 x^{2s} + \alpha^2]\rho) \; . 
\end{equation}
Its steady state solution with reflecting boundary conditions is
\begin{equation} \label{FPss}
\rho(x) = \frac{\rho_o}{1+\lambda^2x^{2s}}
\exp[-\gamma_\alpha \int^x \frac{ y^{r-1}(y-\theta) }{1+\lambda^2y^{2s}}{\rm d}y]  \; ,
\end{equation}
with $\rho_o$ a normalization constant, 
 $\gamma_\alpha\equiv \gamma/\alpha^2$,   
$\lambda^2\equiv (\mu/\alpha)^2\equiv\gamma_\alpha/\gamma_\mu$.
In most cases the integral can be written in terms of 
hypergeometric functions $_2F_1$\cite{abram}, through  
\begin{equation}  \label{int}
\int^x \frac{y^{\beta-1}}{1+\lambda^2y^{2s}}{\rm d}y \equiv
\frac{\beta}{x^\beta}
\,_2F_1(c,1,c+1,-\lambda^2x^{2s})\end{equation} 
with $c\equiv\beta/(2s)\neq -1,-2,\ldots$, 
whereas, in the marginal case $\beta=0$, we will use 
\begin{equation}  \label{intmarg}
\int^x  \frac{y^{-1}}{1+\lambda^2y^{2s}}{\rm d}y\;\equiv\;
\ln x\,-\,\ln(1+\lambda^2x^{2s})/(2s).
\end{equation} 
By means of these definitions and their asymptotic formulas\cite{abram,formula}, 
we obtain  the possible PDF shapes, in $(s,r)$-space, as schematized in Fig.~\ref{fig:am}. 
The marginal cases $r=0$ and $r=-1$ will be considered latter. 
In general, sufficiently large positive $s$ 
is required in order to yield power-law tails,   
otherwise, stretched exponential tails prevail, as for the processes considered in 
Sect. \ref{Sec:mean_rev}. The additive noise does not add 
new domains with power-law tails, although regions with stretched exponential law 
are excluded or included by the normalization condition.  
For vanishing $x$, the main difference with purely multiplicative processes is that, 
for positive both $s$ and $r$, the PDF is truncated at the origin. 
%
Notice that, as the PDF is finite at the origin, then, 
if $x$ is identified with the squared volatility ($x\equiv v^2$), 
the PDF for $v$ increases linearly at the origin.

Let us analyze, in more detail, the marginal cases $r=0$ and $r=-1$  
that can yield power-laws in both asymptotic limits. 
From Eqs.~(\ref{FPss})-(\ref{intmarg}), we obtain   

{\bf A}: For $r=0$, the PDF has the form  
\begin{equation} \label{IIm}
\rho(x) = \rho_o  
\frac{x^{\gamma_\alpha\theta} \Theta(x)}
{[1+\lambda^2x^{2s}]^{\gamma_\alpha\theta/(2s)+1} }  \; , 
\end{equation}
where $\Theta(x)\equiv\exp[-\gamma_\alpha x
\,_2F_1({\scriptstyle \frac{1}{2s},1,\frac{1}{2s}+1},-\lambda^2x^{2s})]$ 
is a smooth function of $x$, 
such that $\Theta(0)$ is finite, hence it does not spoil the power-law growth at the origin. 
For large $x$, it may present different asymptotic behaviors depending on 
the value of $s$:   

{\bf A.1}: If $s\leq 0$, $\Theta(x)$ decays as pure {\em exponential} of $x$. Therefore, the 
asymptotic decay is finally dominated by this exponential factor.

{\bf A.2}: If $0<s<1/2$, $\Theta(x)$ behaves asymptotically as a {\em stretched exponential} 
with argument $x^{1-2s}$.  
That is, the tail, although a power law for moderate $x$, becomes 
asymptotically dominated by a stretched exponential decay. 

{\bf A.3}: If $s>1/2$, $\Theta(x)$ tends to a positive value,  
therefore, in this instance, the tail remains {\em power-law}.

There, by switching $s$, one tunes the tail type, being a power-law for 
$s\ge 1/2$.  In the threshold case $s=1/2$, 
we have $_2F_1(1,1,2,-z)\equiv\ln(1+z)/z$, then we get the explicit expression
\begin{equation} \label{A}
\rho_{\rm A}(x) = \rho_o  
\frac{x^{\gamma_\alpha\theta}}
{[1+\lambda^2x]^{\gamma_\alpha\theta+\gamma_\mu+1} }  \; .
\end{equation}

Thus, the case $r=0$, $s\ge1/2$ allows one to model empirical PDFs with twofold power-law behavior.

{\bf B}: In the case $r=-1$, the normalization condition requires: $s\ge 1/2$, or also, if 
$\gamma_\alpha>1$, $s\le 0$ is allowed. The PDF has the form
\begin{equation} \label{IIIm}
\rho(x) = \rho_o  
\frac{x^{-\gamma_\alpha} \Theta(x)}
{[1+\lambda^2x^{2s}]^{-\gamma_\alpha/(2s)+1} }  \; , 
\end{equation}
where, 
$\Theta(x)\equiv\exp[-\gamma_\alpha\theta\,_2F_1({\scriptstyle -\frac{1}{2s},
1,-\frac{1}{2s}+1},-\lambda^2x^{2s})/x]$ 
tends to a finite value for large $x$, therefore, the tail is a {\em power-law}. 
The asymptotic behavior of $\Theta(x)$ for small $x$, depends on $s$. 

{\bf B.1}: For $s>1/2$, it behaves as an {\em exponential} of $-1/x$, 
that dominates the low $x$ behavior of the PDF. 

{\bf B.2}: For $-1/2<s\le 0$, $\Theta(x)$ behaves as an {\em exponential} of $-1/x^{1+2s}$, that 
dominates the asymptotic behavior. 

{\bf B.3}: However, $\Theta(x)$ takes asymptotically a finite value, if $s<-1/2$; hence, 
the complete expression increases at the 
origin as a {\em power-law}.

At the threshold value $s=-1/2$, by employing again the 
explicit expression for $_2F_1(1,1,2,-z)$, one obtains 
\begin{equation} \label{B}
\rho_{\rm B}(x) = \rho_o\frac{x^{\gamma_\mu\theta+1}}
{[1+x/\lambda^2]^{\gamma_\alpha+\gamma_\mu\theta+1} },  
\;\;\;\;\mbox{if $\gamma_\alpha>1$}\; . 
\end{equation}
Thus, the case $r=-1$, $s\le -1/2$ also provides twofold power-law distributions.

In general,  the class of asymptotic behavior is ruled 
 by $(s,r)$ that determine the form of market laws. This holds, of course, 
as soon as the remaining parameters assume moderate values. 
For instance, the factor $\lambda^2\equiv\gamma_\alpha/\gamma_\mu$ accompanies 
$x^{2s}$ in the formula for $\rho(x)$ [Eqs. (\ref{FPss})-(\ref{intmarg})], 
then, extreme values of $\lambda$ will change the asymptotic regime.  
In fact, in the limit $\alpha=0$ (negligible additive noise, 
corresponding to $\lambda\to \infty$), 
different laws arise, as we have seen in the precedent Section.

Summarizing, we have shown the whole picture of asymptotic behaviors 
that a general class of additive-multiplicative processes produce. 
As a consequence of the extra additive noise, 
new types of asymptotic behaviors emerge. 
Specially interesting solutions arise in the marginal cases $r=0,-1$ 
where two-fold power-law PDFs are found. 

Moreover, additive-multiplicative processes lead to higher richness of crossover 
behaviors, with respect to purely multiplicative processes. 
Therefore, the appearance of new PDF shapes exceeds the one resulting from the 
mere analysis of the asymptotic regimes. 
This is specially important because depending on the values of the parameters, 
the true asymptotic regime might fall outside the observable range.

\section{Comparison with empirical distributions}
\label{Sec:empirical}

Let us consider, as paradigm of the PDFs with two-fold power-law behavior, 
Eqs. (\ref{A}) and (\ref{B}), that have a simple exact expression. 
Actually they have the same functional form, via  
redefinition of parameters ($\theta,\gamma_\alpha,\gamma_\mu$). 
This expression has been recently 
proposed in the literature as an ansatz 
for fitting the distribution of high-frequency  stock-volumes \cite{vol1}, 
under the form 
\begin{equation} \label{fitting} 
\rho(x) = 
\rho_o \frac{(x/x_o)^\nu}
{[1+(q-1)x/x_o]^\frac{1}{q-1} }  \; ,
\end{equation}
where, in that specific application, $x$ is identified with normalized stock volume. 
Therefore, identification of the process for real volumes with one of the 
models above, may allow an econophysical interpretation of the fitting parameters. 
Table I presents the correspondence between the parameters of Eq. (\ref{fitting}) and those  
of processes {\bf A} and {\bf B}, given  by Eqs. (\ref{A}) and (\ref{B}), respectively. 

\vspace*{3mm}
\begin{table}[h]
\begin{tabular}{|c|c|c|} \hline
                             & {\bf A}  & {\bf B}\\
 Eq. (\ref{fitting})         & Eq. (\ref{A}) & Eq. (\ref{B})\\[3mm] \hline\hline
$1/(q-1)$      & $1+\gamma_\alpha\theta+\gamma_\mu$           
& $1+\gamma_\alpha+\gamma_\mu\theta$  \\[3mm] \hline 
$\nu$    & $\gamma_\alpha\theta$       & $1+\gamma_\mu\theta$  \\[3mm] \hline
$x_o$    & $(q-1)\gamma_\mu/\gamma_\alpha$     
   & $(q-1) \gamma_\alpha/\gamma_\mu$ \\[3mm] \hline
\end{tabular}
\caption{Correspondence amongst model parameters.}
\label{table}
\end{table}
\vspace*{3mm}

Recall that $\gamma_\alpha\equiv\gamma/\alpha^2$ and $\gamma_\mu\equiv\gamma/\mu^2$ 
($\lambda^2\equiv\gamma_\alpha/\gamma_\mu$), thus, the power-law exponent for small 
values of $x$, given by $\nu$ (see Table), increases with  
$\gamma$ and $\theta$, and is reduced by either one of the two noise amplitudes: 
the additive noise in process {\bf A} and the multiplicative one in process {\bf B}. 
The power-law decay (with exponent $1/(q-1)-\nu$) for large values of $x$ 
is ruled by either one of 
the effective coefficients $\gamma_\mu$ (in {\bf A}) or $\gamma_\alpha$ (in {\bf B}) 
(see Table). 
That is, the tail is fatter, the larger the corresponding noise amplitude. 
While in process {\bf A} the multiplicative noise affects the tail, in model {\bf B} 
it is affected by the additive noise, oppositely to what happens for small values. 
This is related to the sign of $s$, indicating higher multiplicative 
feedback for either increasing ({\bf A}) or decreasing ({\bf B}) values of $x$. 

\begin{figure}[htb] 
\begin{center} 
\includegraphics*[bb=70 310 510 620, width=0.5\textwidth]{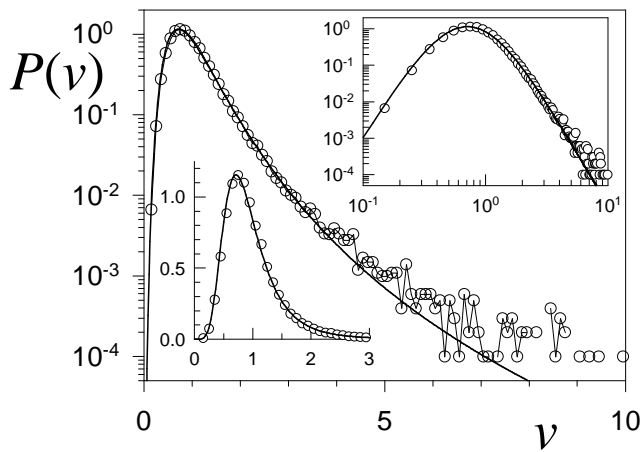} 
\end{center} 
\caption{PDF of normalized volatility of stocks traded in US equity market 
(data from Ref. \cite{micciche}). 
The full line corresponds to a fitting by the theoretical PDF given by 
expression (\ref{fitvol}). Fitting parameters are ($q,\nu,v^2_o)\simeq(1.178,2.20,0.097)$. 
Insets: linear-linear and log-log representation of the same data. } 
\label{fig:data} 
\end{figure}

Besides the good agreement already observed for volumes \cite{vol1,vol2}, 
we tested this functional form to daily data of volatilities reported 
in the literature\cite{micciche}. 
The results are shown in Fig.~\ref{fig:data}. In the models we are generalizing, 
the variable $x$  is usually identified with the variance or squared volatility ($x=v^2$). 
Then, the resulting PDF for $v$ is
\begin{equation} \label{fitvol}
P(v)\;=\;\rho_o \frac{(v/v_o)^{2\nu+1}}
{[1+(q-1)(v/v_o)^2]^\frac{1}{q-1} }  \; ,
\end{equation}
with $\rho_o={\scriptstyle 2(2-q)^\nu(q-1)\Gamma(\frac{1}{q-1}-1)/
[v_o\Gamma(\nu+1)\Gamma(\frac{1}{q-1}-\nu-1)]}$.

Fig.~\ref{fig:data} shows an excellent agreement between theoretical and empirical PDFs, 
for the full range of data. 
Notice that the very central part of the distribution is parabolic in the log-log plot, 
then a poor statistics at the tails may mislead to think that the distribution is 
log-normal. 

\subsection*{Underlying dynamics}

The satisfactory agreement between the empirical data and Eq. (\ref{fitvol}) 
suggests that processes similar to either {\bf A} or {\bf B} may rule  
squared-volatility evolution. 
Hence, let us look at the explicit form of the 
ILEs associated to processes {\bf A} and {\bf B}:
\begin{equation} \label{ILEA}
\mbox{\bf A:}\;\;\;\; {\rm d} x =-\gamma[x-\theta]\frac{1}{x}{\rm d}t + 
\mu \sqrt{x}{\rm d}z + \alpha {\rm d}w \,.
\end{equation}
\begin{equation} \label{ILEB}
\mbox{\bf B:}\;\;\; {\rm d} x =-\gamma[x-\theta]\frac{1}{x^2}{\rm d}t + 
\mu \frac{1}{\sqrt{x}}{\rm d}z + \alpha {\rm d}w \,.
\end{equation}
The first term in each ILE represents the deterministic restoring force with 
respect to the level $\theta$. 
It derives from a confining potential 
of the form  $\gamma[x-\theta \ln x]$ ({\bf A}) or  
$\gamma[\ln x +\theta/x]$  ({\bf B}). 
In both cases, the potential has a minimum located at $x=\theta$ and 
is divergent at $x=0$. 

Average values are 
\begin{equation} \label{averages}
\langle x \rangle_{\bf A} =\frac{\theta+1/\gamma_\alpha}{1-1/\gamma_\mu}
\;\;\;\mbox{and}\;\;\;
\langle x \rangle_{\bf B} =\frac{\theta+2/\gamma_\mu}{1-2/\gamma_\alpha}\;,
\end{equation}
both averages are greater than $\theta$ and coincide only in the limit of 
relatively small noise amplitudes ($\gamma >>\alpha^2,\mu^2$). 
Moments $\langle x^n \rangle$ are finite only if $\gamma_\mu>n$ ({\bf A}) or 
$\gamma_\alpha>n+1$ ({\bf B}). In particular, the second moment is
\begin{equation} \label{s2a}
\langle x^2 \rangle_{\bf A} =\frac{\gamma_\alpha(\theta+1/\gamma_\alpha)(\theta+2/\gamma_\alpha)}
                                  {\gamma_\mu(1-1/\gamma_\mu)(1-2/\gamma_\mu)} \;,
\end{equation}
\begin{equation}
\langle x^2 \rangle_{\bf B} =\frac{\gamma_\mu(\theta+2/\gamma_\mu)(\theta+3/\gamma_\mu)}
                                   {\gamma_\alpha(1-2/\gamma_\alpha)(1-3/\gamma_\alpha)}\;.
\end{equation}
In model {\bf A}, increasing(decreasing) amplitude of the additive(multiplicative) noise, 
increases the width of the 
distribution, whereas model {\bf B} presents opposite behavior. 
Thus, for instance, the additive noise has a confining effect in process {\bf A}, 
opposite to the effect observed in processes with null $\theta$\cite{multi1}.

On the other hand, the distribution has a maximum at 
\begin{equation} \label{maxima}
x^{max}_{\bf A} =\frac{\theta}{1+1/\gamma_\mu}
\;\;\;\mbox{and}\;\;\;
x^{max}_{\bf B} =\theta+\frac{1}{\gamma_\mu}\;.
\end{equation}
Notice that the additive noise does not affect the mode, as expected. 
The most probable value of distribution {\bf A} shifts to the right with increasing 
multiplicative amplitude, while in distribution {\bf B} the opposite tendency occurs.
From Eqs. (\ref{averages}) and (\ref{maxima}),  
$x^{max}_{\bf A}<\theta_{\bf A}<\langle x\rangle_{\bf A}$, 
while $\theta_{\bf B}<x^{max}_{\bf B}<\langle x\rangle_{\bf B}$. 
That is, in model {\bf A}, the reference value $\theta$ represents 
a typical value comprised between two central measures, which does not 
hold in model {\bf B}. This observation, in addition to the positivity of $s$, 
point to model {\bf A} as a more realistic long-term process.

The fitting parameters in Fig.~\ref{fig:data} lead to 
($\theta,\gamma_\mu,\gamma_\alpha)_{\rm \bf A}\simeq(0.50,2.4,4.4)$ or    
($\theta,\gamma_\mu,\gamma_\alpha)_{\rm \bf B}\simeq(0.19,6.3,3.4)$. 
In both cases, $\gamma_\mu,\gamma_\alpha>1$, as expected for regulated markets. 
While $\langle v \rangle =1$, because empirical volatility is normalized, 
$\langle x \rangle =\langle v^2 \rangle \simeq 1.3$ and the mode is $x^{max}\simeq 0.35$, 
consistently with Eqs. (\ref{averages})-(\ref{maxima}). 

Numerical integration of ILEs  (\ref{ILEA})  and (\ref{ILEB}), by 
standard methods\cite{books}, shows that both processes produce  time series 
with bursting or clustering effects, as observed in real sequences.
However, process {\bf B} may present, for some values of the parameters, 
a kind of ergodicity breaking, with large jumps to a state basically 
governed by additive noise.   
This occurs because, once $x$ jumps to a high value, both the restoring force and 
the effective amplitude of multiplicative noise become small as to pull $x$ back to 
its reference level. Then, relaxation is slowed down and the regime of high volatility 
persists for long time stretches. 
Although a process with $s<0$ is not expected to be a realistic model for very long time 
intervals, it can model, for instance, the transient behavior of the market around crashes. 
In fact, process {\bf B} yields akin crises.  
After the crash occurs, this drastic event might 
switch the system back to a $s\ge 0$ regime.

\section{Final remarks}
\label{Sec:final}

We have analyzed stochastic models of a quite general form, 
with algebraic restoring force and algebraic multiplicative noise. 
A further generalization with the inclusion of an extra source of noise, of standard 
Wiener type, has also been analyzed. 
These additive-multiplicative processes are built on the basis of realistic features: 
The multiplicative noise describes innovations generated by endogenous mechanisms that 
amplify or attenuate a random signal, depending on the internal state of the system. 
Whereas, the additive noise encodes a direct influence of external random fields such as news or 
spontaneous fluctuations due to speculative trading. 
One of the goals of this work was to study systematically the PDF asymptotic solutions  
of these generalized models.  
We have shown that the inclusion of additive noise gives rise to new PDF shapes,  
with a richer spectrum of cross-over behaviors and, in particular, two-fold power-law decays. 
The shapes of the PDFs are governed by the effective market rules parametrized by $r$ and $s$. 
These parameters describe the algebraic nature of the global mean-reverting strength of 
the market and the informational coupling among the traders, respectively.
On the other hand, power-law exponents and coefficients of exponential-like functions 
depend also on the reduced parameters $\gamma_\mu$, $\gamma_\alpha$ and on $\theta$. 
This means that one may expect universal behavior among markets that share similar rules 
(same $r$ and $s$) and same rescaled restoring parameters, for a properly normalized 
reference level $\theta$.  
Summarizing, the additive-multiplicative processes given by Eq. (\ref{ILE}) 
provide a general realistic framework to describe the shape of empirical distributions 
for financial, as well as, for physical systems. 
An illustrative application to empirical volatility data was presented 
in Sect.~\ref{Sec:empirical}, showing excellent results.

The statistical description of a market should include its dynamical properties
such as the temporal decay of correlations. 
In real time series of volatilities\cite{liu} and volumes\cite{volumes}, 
power-law decaying correlations have been observed.  
It is worth noting that stochastic processes with additive-multiplicative structure 
(without mean reversion) are  
being currently studied in connection with a generalization of standard (Boltzmann-Gibbs) 
statistical mechanics, recently proposed by C. Tsallis \cite{tsallis}. 
The PDFs associated to this new formalism generalize the exponential weights, 
namely,  $\exp_q(-x) \equiv (1-[1-q]x)^\frac{1}{1-q}$ 
[entering as a factor in Eq. (\ref{fitting})].
The time series arising from additive-multiplicative processes without mean reversion   
present strong correlations that prevent convergence to 
either Gauss or L\'evy limits\cite{multi3} and lead to $q$-Gaussian distributions. 
This suggests that similar correlations may persist in mean-reverting processes 
with the additive-multiplicative character.  
Once Eq. (\ref{ILE}) leads to PDFs in such a good agreement with empirical ones, it is 
worth performing a detailed study and comparison of real and artificial time series 
to test the models with respect to the dynamics. 
Elucidating this point deserves a careful separate treatment.

{\bf Acknowledgments: }
We are grateful to S. Miccich\`e, G. Bonanno, F. Lillo and R.N. Mantegna 
for communicating their numerical data in Ref. \cite{micciche}.


\begin{thebibliography}{99}
\bibitem{formula} The hypergeometric function $f(z)=$$_2F_1(c,1,c+1,-y)$  
 behaves asymptotically as  $f(y)\sim 1$,  
for small $y$, and  as $(1/y,\ln[1+y]/y,1/y^c$), if $c(>,=,<)1$, for large $y$.
\end{thebibliography}
\end{document}